# Phononic frequency comb via three-mode parametric three-wave mixing

Authors: Adarsh Ganesan[1], Cuong Do[1], Ashwin Seshia[1]

[1.] Nanoscience Centre, University of Cambridge, Cambridge, UK

**This paper is motivated by the recent demonstration of three-wave mixing based phononic frequency comb. While the previous experiments have shown the existence of three-wave mixing pathway in a system of two-coupled phonon modes, this work demonstrates a similar pathway in a system of three-coupled phonon modes. The paper also presents a number of interesting experimental facts concomitant to the three-mode three-wave mixing based frequency comb observed in a specific micromechanical device. The experimental validation of three-mode three-wave mixing along with the previous demonstration of two-mode three-wave mixing points to the ultimate possibility of multimode frequency combs.**

Optical frequency combs have significantly transformed modern metrology and molecular spectroscopy [1-2]. Recently, we experimentally demonstrated the existence of such frequency combs in the phononic domain [3] after its theoretical prediction in [4]. While both optical and phononic frequency combs carry similar spectral features, the dynamics describing the respective generation processes are different. The optical frequency combs usually arise through a well-established Kerr nonlinear pathway [4]. However, the pathway describing our more recent phononic frequency combs is nonlinear three-wave mixing [4]. Through this mechanism, a single drive tone intrinsically couples with the eigenfrequency of a phonon mode. Such physical interaction can particularly enable high-precision micro and nano-mechanical resonant sensors adapted for the direct monitoring of slow varying intrinsic or extrinsic physical processes.

In the phononic frequency comb demonstrated in [3], the three-wave mixing process is particularly operative in the system of two parametrically coupled phonon modes. However, the preceding theoretical work [4] on frequency combs have actually shown the possibility for frequency combs in a system comprising three and four parametrically coupled phonon modes. Hence, inspired by these mathematical predictions [4], we now experimentally describe three wave mixing process even in three mode parametric resonance [5]. Additionally, this paper also presents surprising experimental facts specific to three-mode three-wave mixing based phononic frequency comb.

For understanding the surprising nature of three-mode three-wave mixing based phononic frequency comb, we first present a brief discussion of two-mode three-wave mixing based phononic frequency comb. Hence, we first consider the coupled two-mode dynamics.

$$\ddot{Q}_i = -\omega_i^2 Q_i - 2\zeta_i\omega_i \dot{Q}_i + \sum_{\tau_1=1}^{2}\sum_{\tau_2=1}^{2} \alpha_{\tau_1\tau_2} Q_{\tau_1}Q_{\tau_2} + \sum_{\tau_1=1}^{2}\sum_{\tau_2=1}^{2}\sum_{\tau_3=1}^{2} \beta_{\tau_1\tau_2\tau_3} Q_{\tau_1}Q_{\tau_2}Q_{\tau_3} + P\cos(\omega_d t) \tag{1}$$

where $P$ is the drive level, $\alpha$ and $\beta$ are quadratic coupling coefficients and $\omega_{i=1,2}$ and $\zeta_{i=1,2}$ are natural frequencies and damping coefficients of modes $i=1,2$ respectively.

Here, when the drive frequency $\omega_d$ is closer to $\omega_1$, the direct excitation of mode 1 takes place. The larger drive level $P$ cause higher displacement $Q_1$ which in turn results in the parametric excitation of mode 2 through the nonlinear term $Q_1Q_2$. The frequency of this parametrically excited tone is expected to be $\frac{\omega_d}{2}$ in the case of two-mode parametric resonance [6]. However, owing to an intrinsic three-wave mixing pathway, the excitation of $\frac{\omega_1}{2}$ tone instead of $\frac{\omega_d}{2}$ is observed in our recent demonstration of phononic frequency comb. Also, this specific three-wave mixing pathway is only operative when $\omega_d$ is set outside the dispersion band of driven phonon mode. Subsequent to this excitation, through high-order interactions, a frequency comb of spacing $|\omega_d - \omega_1|$ is formed about $\omega_1$ and $\frac{\omega_1}{2}$.

While the concept of three-wave mixing with two-mode parametric resonance is clear, we now turn to three-wave mixing with three-mode parametric resonance. For understanding this, we consider a system of three coupled modes [5].

$$\ddot{Q}_i = -\omega_i^2 Q_i - 2\zeta_i\omega_i \dot{Q}_i + \sum_{\tau_1=1}^{3}\sum_{\tau_2=1}^{3} \alpha_{\tau_1\tau_2} Q_{\tau_1}Q_{\tau_2} + \sum_{\tau_1=1}^{3}\sum_{\tau_2=1}^{3}\sum_{\tau_3=1}^{3} \beta_{\tau_1\tau_2\tau_3} Q_{\tau_1}Q_{\tau_2}Q_{\tau_3} + P\cos(\omega_d t); \; i=1,2,3 \tag{1}$$

Based on this dynamics, for $\omega_d \cong \omega_1 \cong (\omega_1 + \omega_2)$, the parametric excitation of tones $\omega_x \cong \omega_1$ and $\omega_y \cong \omega_2$ which satisfies the condition $\omega_x + \omega_y = \omega_d$ is expected. Using the perspective derived from two-mode three-wave mixing, we now outline the three-wave mixing pathway in this system of three-coupled modes.

In the case of two-mode three-wave mixing, one of the frequencies corresponding to the combs of spacing $|\omega_d - \omega_1|$ had to satisfy the condition $2\omega = \omega_1$ instead of $2\omega = \omega_d$. Hence, a frequency comb of $\frac{\omega_1}{2} \pm n(\omega_d - \omega_1)$ is formed instead of $\frac{\omega_d}{2} \pm n(\omega_d - \omega_1)$. On similar lines, in the case of three-mode three-wave mixing, one might expect to have two frequencies $\omega_p$ and $\omega_q$ the respective frequency combs of mode 2 and 3 that satisfy a frequency matching condition of $\omega_p + \omega_q = \omega_1$

instead of $\omega_p + \omega_q = \omega_d$. In the frequency combs: $\omega_x \pm n(\omega_d - \omega_1)$; $\omega_y \pm n(\omega_d - \omega_1)$, such a condition of $\omega_p + \omega_q = \omega_1$ is satisfied when $\omega_p = \omega_x$; $\omega_q = \omega_y - (\omega_d - \omega_1)$. Additionally, unlike two-mode three-wave mixing, the frequency condition of $\omega_p + \omega_q = \omega_d$ is also satisfied in these frequency combs when $\omega_p = \omega_x$; $\omega_q = \omega_y$.

In order to establish this pathway for comb formation, we experimentally probe a micromechanical device and organise the experimental observations to support the above discussion. Note: In the experimental section, unlike in the theory section, the temporal frequency $f$ is used in the discussion instead of angular frequency $\omega = 2\pi f$.

For studying the three-mode three-wave mixing based frequency comb, the same experimental system that was used to study two-mode three-wave mixing [3] is once again considered i.e. an AlN-on-Si free-free micro-beam of dimensions $1100 \times 350 \times 11 \ \mu m^3$ (Figure 1A). A sinusoidal electrical signal is applied through one of the split electrodes patterned on the microstructure and the output signal which is extracted via another split electrode is analysed using Agilent infiniium 54830B DSO. The experiments were carried out under ambient pressure and temperature conditions.

Figure 1B shows the frequency spectrum of output electrical signal when $S_{in}(f_d = 3.857 \ MHz) = 15 \ dBm$ is applied. In this spectrum, we can see five thick spectral features b1-b5. The frequency corresponding to b1 is close to the drive frequency $f_d \cong f_1$. While the features b2 and b3 are associated with frequencies $f_m \cong f_2$ and $f_n \cong f_3$ respectively, their sum is seen to approximately equal to the drive frequency. The final two features b4 and b5 have frequencies $2f_m$ and $2f_n$ respectively and these correspond to the second harmonics of features b2 and b3 respectively. To clearly visualize each of these spectral features, the zoomed-in images are presented in the figures 1b1-1b5. These figures clearly show that b1-b5 correspond to the frequency combs of spacing $5.035 \ kHz$ about the frequencies $3.857 \ MHz = f_d$, $1.791 \ MHz \cong f_2$, $2.066 \ MHz \cong f_3$, $3.582 \ MHz \cong 2f_2$ and $4.132 \ MHz \cong 2f_3$ respectively. The tone corresponding to the drive frequency $f_d$ can be located in the figure 1b1 and the additional spectral lines in the output spectrum arise through the nonlinear three-wave mixing process.

To systematically understand the evolution of three-mode three-wave mixing based frequency comb, the experiments are carried out for a range of drive levels: $4 - 23.8 \ dBm$ and the frequency combs about $f_d \cong f_1$, $f_m \cong f_2$ and $f_n \cong f_3$ are examined. Figure 2 shows the drive level dependence of the frequency combs. Each of the frequency combs about $f_d$, $f_m$ and $f_n$ presented in figures 2A-2C correspond to different phonon modes and their mode shapes are presented in figure 2 inset. The vertical line in figure 2A corresponds to $f_d$ and the additional lines are formed about $f_d$ with equidistant spacing. Interestingly, this spacing is found to increase with the drive level $S_{in}$. Additionally, at higher drive levels $S_{in} \geq 18\ dBm$, inter-leaved spectral lines are also formed. This is broadly similar to the case presented in our previous demonstration of three-wave mixing [3]. Corresponding to the frequency generation about $f_d$, combs are also formed about $f_m \cong f_2$ and $f_n \cong f_3$ (Figure 2B and 2C). However, there are no vertical lines in figures 2B and 2C. This shows that the frequencies $f_m$ and $f_n$ corresponding to the two parametrically excited internal modes are also drive level dependent and such dependences are related to the nonlinear comb generation process. This leads to a drive-level dependence of the comb spacing in the presented pathway.

To understand more about the drive level dependence of the frequency comb, the evolution of frequencies $f_m$ and $f_n$ is investigated. It can be seen from figures 3A and 3B that $f_m$ and $f_n$ are not simply drive level dependent but their dependences are also characterized by peculiar nonlinear functions. Despite this, $f_m + f_n$ is always equal to $f_d$ for any drive level $S_{in}$. The experimental facts presented in the figures 3A-3C also confirm the specific nature of frequency combs observed in our device is $(f_m = f_x) \pm n(f_d - f_1); (f_n = f_y) \pm n(f_d - f_1); f_d \pm n(f_d - f_1)$. This frequency comb possesses an equidistant spacing of $|f_d - f_1|$. The figure 3D shows that the resonant frequency $f_1$ is drive level dependent although linear. This is similar to the case observed in [3]. This drive level dependence of $f_1$ thus leads to the drive level dependent comb spacing as noted in figure 2A. Figure 4A further validates that the spacing of frequency combs formed about $f_d$, $f_x$ and $f_y$ are equal. To obtain the relevance of the dispersion band of the driven phonon mode on the frequency comb, the experiments were conducted at different drive frequencies. Despite the significant low 3-mode parametric resonance threshold within the dispersion band for instance $f_d = 3.856\ MHz$, the frequency comb is only existent outside the dispersion band specifically on the right side (Figure 4B). The reason for the existence of an asymmetric frequency comb on only one side of dispersion band, unlike the case presented in [3], can possibly be explained by the asymmetry of drive frequency dependent parametric excitation threshold (Figure 4B). In the frequency combs observed in our experiments, similar to those presented in [3], the spacing stays the same for different drive frequencies at a specific drive level in addition to the increase in spacing with drive level for a specific drive frequency. This can also be evidenced from the colour-maps in the figure 4B.

Now, we turn to the interesting trend presented in the figures 3A-3C. Frequencies $f_x$ and $f_y$ shift symmetrically while maintaining the condition $f_x + f_y = f_d$. The magnitude of this shift possesses a peculiar nonlinear relationship with drive level. To prove the relevance of three-mode three-wave mixing process on this trend, the nature of drive level dependence of $f_x$ and $f_y$ for different drive frequencies are examined. It is possible to clearly note from the figures 5A and 5B that the simultaneous downshifts and upshifts in $f_x$ and $f_y$ respectively are only observed when the three-mode three-wave mixing is existent. Despite these shifts, $f_x + f_y$ for all drive levels and drive frequencies are equal to the respective $f_d$. Further, for the drive frequencies that constitute three-mode three-wave mixing, the respective nonlinear relationships with drive levels are both different and non-monotonous. Such relationships may arise through the nonlinear feedback involved in the frequency comb generation process and may also offer an additional room for rigorous fundamental establishment despite that they fall outside the scope of this current manuscript. In addition to the mountains and pits of the respective figures 5A and 5B, we can also observe another interesting characteristic associated with the three-mode three-wave mixing. Even in the absence of three-mode three-wave mixing, there exists an obscure nominal relationship between $f_x$ & $f_y$ and $f_d$ which is clear from the planes presented in the figures 5A-5C and any influence of three-wave mixing is only above these nominal planes. These planes are parallel to the drive level axis as the frequencies $f_x$ and $f_y$ for a specific $f_d$ do not vary with the drive level and such possible drive level dependences are only observed under the influence of three-wave mixing. Figures 5A and 5B show that the slopes of respective planes are not same. This suggests that $f_y - f_x$ is not constant with $f_d$ and such specific relationship along with the comb induced symmetric frequency shifts should also be theoretically understood.

This paper thus presents the first experimental demonstration of a phononic frequency comb via three-mode three-wave mixing using a micromechanical resonator. The specific experimental facts associated with this form of frequency comb are also provided. The validated existence of phononic frequency combs via both two-mode three-wave mixing and three-mode three-wave mixing can thus create a general perceptive for multi-mode three-wave mixing based frequency combs. Such multi-mode frequency combs can possibly be helpful in the distribution of frequency combs to the multiple segments of frequency spectrum.

**Acknowledgements**

Funding from the Cambridge Trusts is gratefully acknowledged.

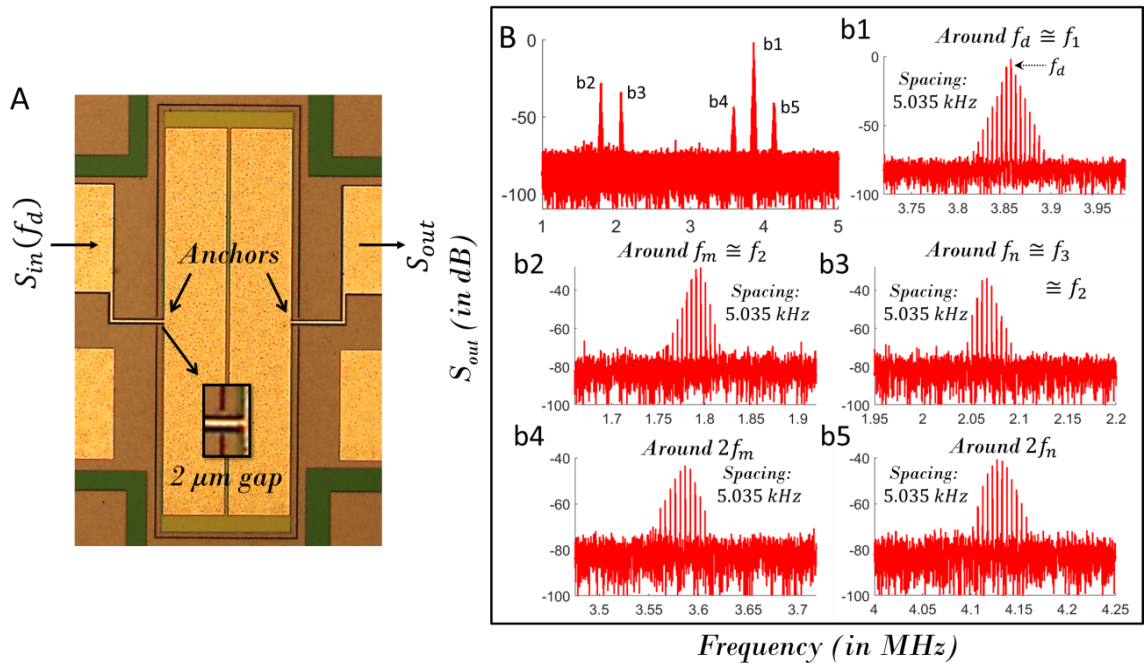

Figure 1: **Observation of phononic frequency comb via three-mode three-wave mixing.** A: An electrical signal $S_{in}(f_d = 3.857\ MHz)$ is provided to a free-free beam microstructure; B: The frequency spectrum of the output electrical signal $S_{out}$; b1-b5: The zoomed views of spectral features b1-b5 in B respectively.

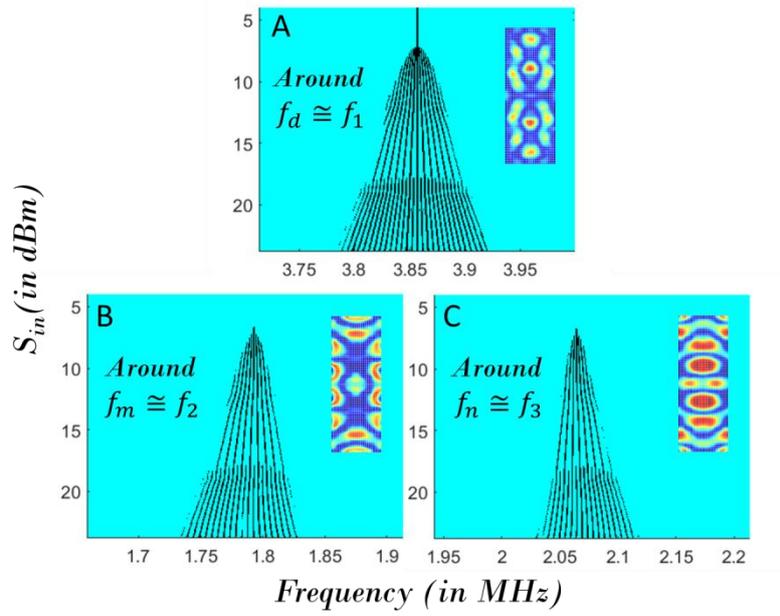

Figure 2: **Drive level dependence of frequency comb.** A-C: The spectral maps of output electrical signal $S_{out}$ around $f_d$, $f_m$ and $f_n$ respectively for different drive conditions $S_{in}(f_{d1} = 3.86\, MHz) = 4 - 23.8\, dBm$. The inset figures show the vibration mode shapes corresponding to the respective frequency combs and the red and blue correspond to maximum and minimum displacements in these inset figures.

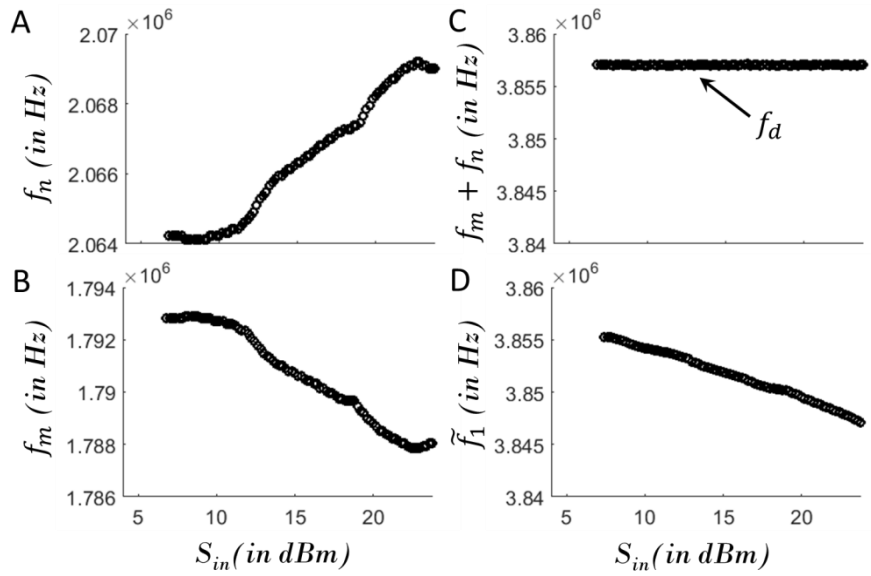

Figure 3: **Drive level dependence of frequency comb (Contd.).** A-D: The drive level $S_{in}$ dependence of $f_n$, $f_m$, $f_m + f_n$ and $\tilde{f}_1$ respectively.

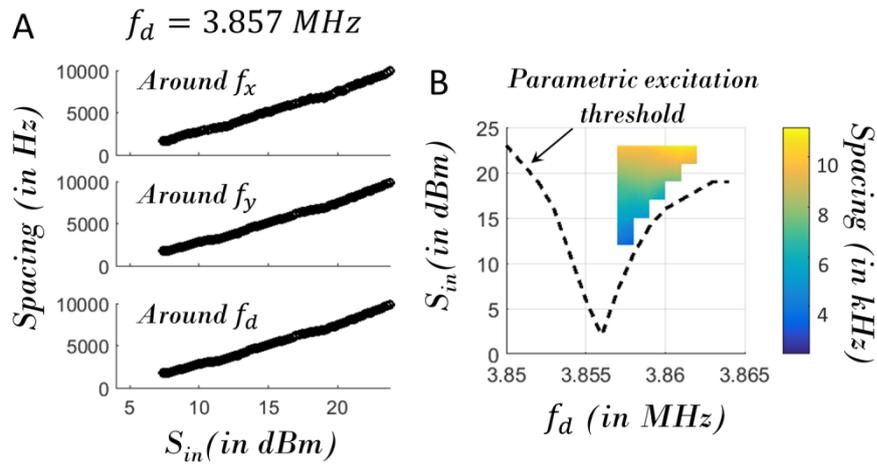

Figure 4: **Drive frequency dependence of frequency comb.** A: The spacing of frequency combs around $f_x$, $f_y$ and $f_d$ for the drive frequency $f_d = 3.857\ MHz$; B: The spacing of frequency combs for different drive frequencies $f_d$ and drive levels $S_{in}$. The colour-maps indicate the spacing. The absence of colour or white-colour indicates the absence of frequency comb for that drive condition. The dotted black line indicates the parametric excitation threshold. The drive level $S_{in}$ above this threshold line leads to parametric resonance.

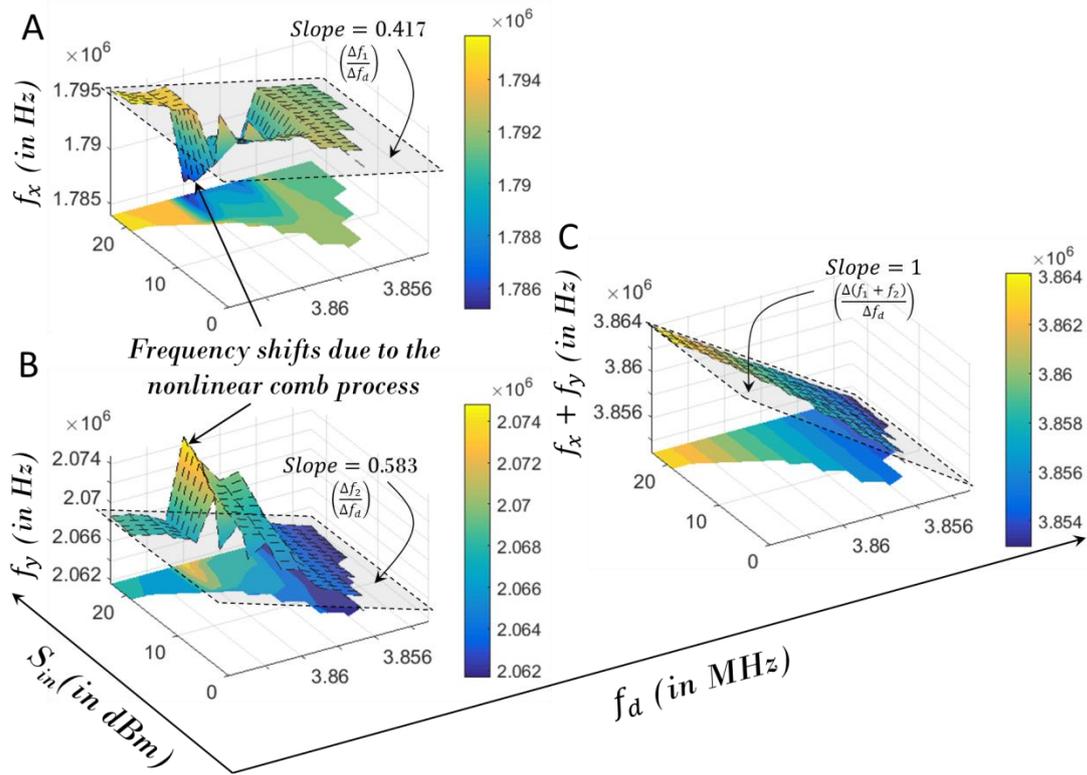

Figure 5: **Drive frequency dependence of frequency comb (Contd.).** A-C: The value of $f_x$, $f_y$ and $f_x + f_y$ for different drive frequencies $f_d$ and drive levels $S_{in}$ respectively. The colour-maps indicate the values of these frequencies. The absence of colour or white-colour indicates the absence of frequency comb for that drive condition. The sketched planes correspond to the nominal drive frequency dependence of $f_x$, $f_y$ and $f_x + f_y$ i.e. under the absence of three-wave mixing. Note: The projections of 3-D plots on the $S_{in} - f_d$ plane are also shown for clarity.